# Riddling and chaotic synchronization of coupled piecewise-linear Lorenz maps

M. C. Vergès, R. F. Pereira, S. R. Lopes, R. L. Viana *

*Departamento de Física, Universidade Federal do Paraná,*

*Caixa Postal 19044, 81531-990, Curitiba, Paraná, Brazil*

T. Kapitaniak

*Division of Dynamics, Technical University of Lódz,*

*Stefanowskiego 1/15, 90-924 Lódz, Poland*

(Dated: February 9, 2009)

## Abstract

We investigate the parametric evolution of riddled basins related to synchronization of chaos in two coupled piecewise-linear Lorenz maps. Riddling means that the basin of the synchronized attractor is shown to be riddled with holes belonging to another basin in an arbitrarily fine scale, what has serious consequences on the predictability of the final state for such a coupled system. We found that there are wide parameter intervals for which two piecewise-linear Lorenz maps exhibit riddled basins (globally or locally), what indicates that there are riddled basins in coupled Lorenz equations, as previously suggested by numerical experiments. The use of piecewise-linear maps makes it possible to prove rigorously the mathematical requirements for the existence of riddled basins.

---

* Corresponding author. e-mail: viana@fisica.ufpr.br

## I. INTRODUCTION

In 1963 Lorenz published a landmark paper which marked the beginning of the modern age of nonlinear dynamics [1]. One of the most striking insights of Lorenz was his perception that the complicated (in fact, chaotic) dynamics exhibited by the three-dimensional flow for some parameter values could be understood in terms of a simpler, low-dimensional discrete map. This was obviously a consequence of the strong dissipative character of the equations, and it is a widely used procedure in modern nonlinear dynamics research. Lorenz took the local maxima of one of the evolving variables of his system and plotted a first return map, the so-called Lorenz map. It turned out that the form of this map resembled the so-called tent function ($x \mapsto 1 - 2|x - 1/2|$) whose properties were already known by mathematicians for a long time [2].

This resemblance between the Lorenz and tent maps was serendipitous since the latter was known to possess ergodic orbits (now called transitive or strongly chaotic [2]), such that it provided a sound foundation for the claim that the Lorenz original system would present chaotic trajectories in the phase space. Moreover, the correspondence between the butterfly attractor and the properties of the Lorenz map furnished an early evidence that chaotic dynamics in complex, higher-dimensional systems could be understood by studying simpler, low-dimensional systems.

The Lorenz map is non-smooth thanks to its having a cusp, but it is smooth (albeit nonlinear) otherwise. On the other hand, we may have some advantages of using a piecewise-linear Lorenz map, also known as the bungalow-tent map [3, 4]. Piecewise-linear maps have been intensively studied as models of a variety of physical systems ranging from electrical circuits [5] to vibroimpact mechanical oscillators [6]. Moreover, chaotic piecewise-linear maps have been shown to be *robust* in the sense that the chaotic orbits would persist for parameter variations, a property not shared by smooth chaotic systems in general [5].

In this paper we will explore some dynamical properties of two coupled piecewise-linear Lorenz maps. This is a simple system which nonetheless contains some of the fundamental properties of coupled map lattices and even oscillator chains. Being piecewise-linear such coupled maps have the additional advantage of having a constant Jacobian and thus some analytical exact results for, e.g. Lyapunov spectra [7]. This allows one to investigate phenomena like complete synchronization of chaos and riddled basins [8, 9].

In brief, a riddled basin is a basin of attraction of an attractor characterized by the presence of holes belonging to the basin of another attractor [10, 11]. As a consequence, given an initial condition belonging to the basin of some attractor, there will be always some points in its neighborhood, no matter how small it is, belonging to the basin of another attractor. Hence the physical consequences of riddling are quite serious in terms of our ability of predicting what attractor the trajectory originating from a given initial conditions asymptotes to.

Kim and coworkers [12] have suggested the presence of riddled basins in two diffusively coupled Lorenz equations, what comprises a six-dimensional system where the synchronized orbit lies on a three-dimensional subspace. However, a direct characterization of riddling is quite difficult in such high-dimensional systems, calling for indirect ways to evidence riddling. One example is to consider the coupling of Lorenz maps, which can give us insight on the mechanisms whereby two coupled Lorenz systems can generate riddled basins. However, since the original Lorenz map would be likewise difficult to analyze, a better approach could be to use a piecewise linear approximation to it, as we do in this paper.

Accordingly, our aim in this work is to describe the parametric evolution of riddled basins of chaotic synchronization in two coupled piecewise-linear Lorenz maps. We shall describe with particular emphasis two situations the parameter ranges for which the coupled maps present riddled basins. The parameters to be considered are the slope of the isolated maps and the coupling strength, considering two different coupling prescriptions. We show that the properties of riddled basins are qualitatively distinct according to the prescription used.

The rest of the paper is organized as follows: in the second Section we outline the properties of the piecewise-linear Lorenz map. Section III deals with the synchronization prescriptions for two coupled maps, with some exact analytical results. The study of riddled basins of synchronization are treated in Section IV. The last Section contains our Conclusions.

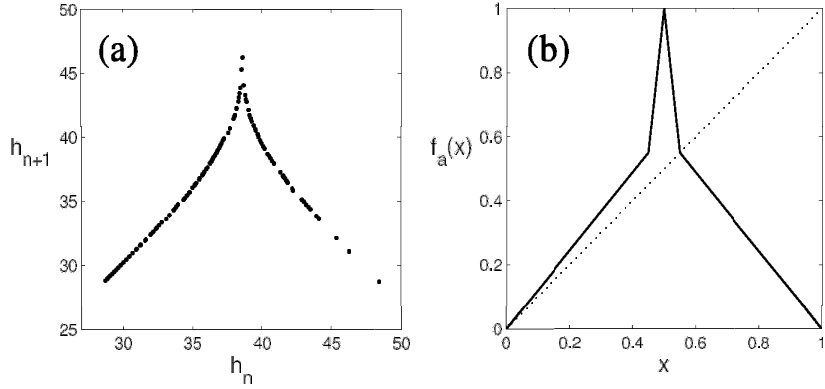

FIG. 1: (a) Lorenz map. (b) Bungalow-tent map for $a = 0.45$.

## II. THE PIECEWISE-LINEAR LORENZ MAP

It is a well-known fact that the Lorenz equations,

$$\frac{dx}{dt} = \alpha(y - x), \tag{1}$$

$$\frac{dy}{dt} = x(\beta - z) - y, \tag{2}$$

$$\frac{dz}{dt} = xy - \gamma z \tag{3}$$

produce phase-space trajectories which converge to a chaotic attractor with an involved fractal structure when $\alpha = 10$, $\beta = 28$, and $\gamma = 8/3$ [1]. One way to convey this idea is to perform a dimensional reduction by sampling only the local maxima of one of the phase-space variables, so as to get a discrete-time variable $h_n = \max_{t=t_n}\{z(t)\}$. Then we build a first return plot by graphing $h_{n+1}$ versus $h_n$, the result being the so-called Lorenz map $h_{n+1} = L(h_n)$, which is a unimodal map with a cusp [Fig. 1(a)]. Such maps display a similar dynamical behavior as the tent map ($x_{n+1} = 1 - 2|x_n - 1/2|$) and, in fact, this similarity was used by Lorenz to explain the erratic behavior at the chaotic attractor in the three-dimensional phase space [1].

Another piecewise-linear approximation to the Lorenz map is provided by the so-called bungalow-tent map, which has out of four linear segments instead of two, as in the tent

function [3, 4]:

$$x_{n+1} = f_a(x_n) = \begin{cases} \frac{1-a}{a} x_n & \text{if } x_n \in [0, a) \\ \frac{2a}{1-2a} x_n + \frac{1-3a}{1-2a} & \text{if } x_n \in [a, \frac{1}{2}) \\ \frac{2a}{1-2a}(1-x_n) + \frac{1-3a}{1-2a} & \text{if } x_n \in [\frac{1}{2}, 1-a) \\ \frac{1-a}{a}(1-x_n) & \text{if } x_n \in [1-a, 1] \end{cases} \quad (4)$$

where $a \in (0, \frac{1}{2})$ is a control parameter. For $a = \frac{1}{3}$ we obtain the tent map. A linear coordinate transformation $h = 22x + 32$ and the choice $a = 0.45$ would approximate the Lorenz map better than the tent function [Fig. 1(b)]. In the following we will refer to the former as the piecewise-linear Lorenz map.

The two fixed points of the bungalow map are $P_1 : x_1^* = 0$ and $P_2 : x_2^* = 1 - a$. For all values of the control parameter $a$, the fixed points are unstable because their eigenvalues satisfy $|\delta_{1,2} = (a-1)/a| > 1$. Although the map is non-smooth at the fixed point $x_2^*$, we can define the map derivative at the right-hand site of $x_2^*$, because this point belongs to the interval $[1 - a, 1]$.

A useful feature of the bungalow map is that the Lyapunov exponent is a smooth function of the control parameter $a$. In order to obtain this dependence we solve the Frobenius-Perron equation

$$\rho_a(x) = \int_0^1 \rho_a(y) \, \delta(x - f_a(y)) \, dy. \quad (5)$$

to obtain the invariant density [3]

$$\rho_a(x) = \frac{1}{2 - 3a} \chi_{[0, 1-a)}(x) + \frac{1 - 2a}{a(2 - 3a)} \chi_{[1-a, 1]}(x), \quad (6)$$

where $\chi$ is the indicator function, i.e., $\chi_A(x) = 1$ if $x \in A$, and $\chi_A(x) = 0$ if $x \notin A$. Thus the invariant density is constant in the interval $[0, 1 - a)$ and jumps, at $x = 1 - a$, to another constant value. The Lyapunov exponent is then calculated using

$$\begin{aligned} \lambda(a) &= \int_0^1 \rho_a(x) \ln \left| \frac{df_a(x)}{dx} \right| dx. \\ &= \frac{1-a}{2-3a} \ln\left(\frac{1-a}{a}\right) + \frac{1-2a}{2-3a} \ln\left(\frac{2a}{1-2a}\right). \end{aligned} \quad (7)$$

In particular, for $a = 1/3$ we obtain $\lambda(1/3) = \ln 2$, which is the expected result for the tent map [2], and also turns to be the maximum possible value for this exponent. For other values of $a$, the Lyapunov exponent is always positive, displaying a smooth dependence with $a$ [Fig. 2], and assuming the value $\lambda = 0.5078$ for the approximation to the Lorenz map which results from $a = 0.45$. Hence we shall always be treating a chaotic (actually a transitive) map, with no periodic windows irrespective of the value of the parameter $a$.

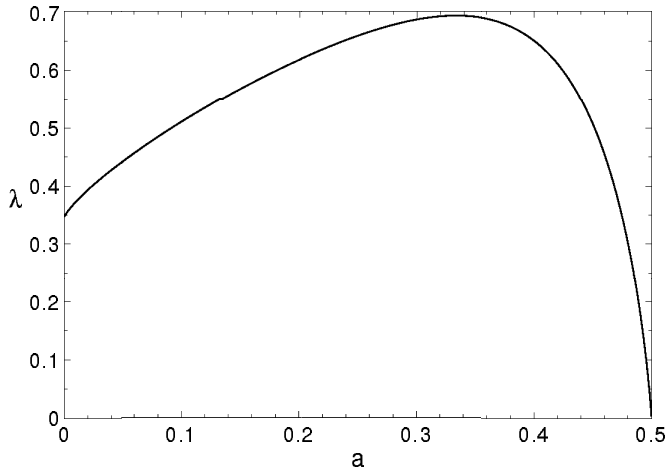

FIG. 2: Lyapunov exponent of a piecewise-linear Lorenz map *versus* its parameter $a$.

## III. SYNCHRONIZATION OF TWO COUPLED PIECEWISE-LINEAR LORENZ MAPS

Coupled maps are paradigmatic examples of higher-dimensional dynamical systems, and have been extensively used since the 80s to investigate spatially extended systems where there is some kind of interplay between temporal and spatial degrees of freedom. One of such phenomena is synchronization of chaos [13]. Coupled chaotic maps can synchronize their amplitudes at each time depending on the coupling strength. The synchronized state is a low-dimensional attractor embedded in the system phase space, and its stability properties have been studied with respect to perturbations along transversal directions. Most of these investigations, however, are chiefly numerical since the system Jacobian depends on the phase space variables. Hence much effort has been devoted to coupled piecewise linear maps, for which the Jacobian matrix has constant entries, what simplifies enormously the calculations and shed light on the mechanism whereby the synchronized state loses transversal stability [9].

The simplest problem belonging to this category consists of two coupled piecewise-linear

Lorenz maps. Remembering that the Lorenz system describes a temporal dynamics for a fixed spatial location (the spatial structure being assumed *a priori* as given), the system of coupled maps could be regarded as a rough model of what happens if we consider the time evolution at two different spatial locations in the forced convection experiment from which we derive the Lorenz equations. In this case a natural choice for the coupling function is the diffusive one, i.e. it depends on the difference between the variables for both maps.

As a matter of fact, Kim *et al.* [12] have considered two identical Lorenz systems diffusively coupled in their $z$-component, or

$$\frac{dx_i}{dt} = \alpha(y_i - x_i), \tag{8}$$

$$\frac{dy_i}{dt} = x_i(\beta - z_i) - y_i, \tag{9}$$

$$\frac{dz_i}{dt} = x_i y_i - \gamma z_i + \epsilon(z_{i+1} - z_i) \tag{10}$$

with $i = 1, 2$, such that $z_3 \equiv z_1$ and so on, $\epsilon$ being the coupling strength. This system was found to present both a synchronized and an anti-synchronized state, defined, respectively, by

$$(x_1 = x_2, y_1 = y_2, z_1 = z_2), \quad (x_1 = -x_2, y_1 = -y_2, z_1 = z_2),$$

which are contained in three-dimensional invariant subspaces of the six-dimensional phase space of the coupled system. There have been found numerical evidences of riddled basins for values of $\epsilon$ large enough, but a systematic investigation is quite difficult in view of the high dimensionality of the system. Hence we can resort to the one-dimensional reduction performed by taking the Lorenz map and its piecewise-linear counterpart.

There are two different forms of coupling two piecewise-linear Lorenz maps, what renders two distinct two-dimensional maps, to be referred to as $\mathbf{F}$ and $\mathbf{G}$ given, respectively, by

$$\text{Version 1:} \begin{pmatrix} x_{n+1} \\ y_{n+1} \end{pmatrix} = \mathbf{F}\begin{pmatrix} x_n \\ y_n \end{pmatrix} = \begin{pmatrix} f_a(x_n) + \delta(y_n - x_n) \\ f_a(y_n) + \varepsilon(x_n - y_n) \end{pmatrix} \tag{11}$$

$$\text{Version 2:} \begin{pmatrix} x_{n+1} \\ y_{n+1} \end{pmatrix} = \mathbf{G}\begin{pmatrix} x_n \\ y_n \end{pmatrix} = \begin{pmatrix} f_a(x_n + \delta(y_n - x_n)) \\ f_a(y_n + \varepsilon(x_n - y_n)) \end{pmatrix} \tag{12}$$

where $\delta$ and $\varepsilon$ are coupling strengths which can be different when the coupling is asymmetric, and even vanish for unidirectional coupling (a *master-slave* configuration). As it will turn out, the dynamical behavior of the system depends essentially on their sum $d \equiv \delta + \varepsilon$.

The basic difference between these kinds of maps is that $\mathbf{F}$ does not render the square $[0,1]^2$ invariant under its action, unless both $\delta$ and $\varepsilon$ are zero, whereas $\mathbf{G}$ keeps the square invariant when both $0 \leq \delta \leq 1$ and $0 \leq \varepsilon \leq 1$. In fact, in order to always have **F** and **G** as well-defined transformations of $\mathbb{R}^2$, we must extend the domain of the map variables to the whole of $\mathbb{R}$ by omitting the restrictions $0 \leq x$ and $x \leq 1$ in (4). In any case, the segment $\mathcal{S} = \{(x, y) \mid 0 \leq x = y \leq 1\}$ is kept invariant and is a subspace where the two coupled maps $x_{n+1} = f_a(x_n)$ and $y_{n+1} = f_a(y_n)$ are synchronized. It will be referred to as the *synchronization subspace* from now on. The orbit in the synchronization subspace is obviously the same as a chaotic orbit from the uncoupled map, hence the fixed points in $\mathcal{S}$ are also $P_1 = 0$ and $P_2 = 1 - a$, located at the intervals 1: $[0, a)$ and 4: $[1 - a, 1]$, respectively.

In order to check this stability of the synchronization subspace we have to investigate whether or not a trajectory that starts slightly off $S$ converges to a trajectory in $S$. This can be done by considering the maximum Lyapunov exponent along the direction transversal to the synchronization subspace, denoted as $\lambda_\perp$. If it is negative (positive) the synchronized state is stable (unstable) with respect to infinitesimal transversal perturbations.

This is only a necessary condition, though, since there may be transversely unstable periodic orbits embedded in the chaotic orbit which lies on the synchronization subspace. Such transversely unstable orbits may exist even when $\mathcal{S}$ is transversely stable (i.e. with $\lambda_\perp < 0$), and their overall effect is to drive trajectories away from $\mathcal{S}$, spoiling the convergence to the synchronized state. Hence a sufficient condition for a stable synchronized orbit is that all unstable periodic orbits in the synchronized state be transversely stable. This last requirement can be verified by computing the eigenvalue of the Jacobian matrix along the transversal directions such that all of them have moduli less than the unity.

### A. Version I

Let us now apply these general arguments to the particular case of the coupled piecewise-linear Lorenz maps. Introducing the variables $u = x - y$ and $v = x + y$ there results that the synchronization manifold $\mathcal{S}$ is given by $u = 0$, and the transversal direction is parameterized by the $v$-coordinate. We first consider the coupled system in its version I [Eq. (11)], for which the Jacobian matrix is

$$\mathbf{DF} = \begin{pmatrix} c - \delta & \delta \\ \varepsilon & c - \varepsilon \end{pmatrix}, \tag{13}$$

where

$$c = \begin{cases} \frac{1-a}{a} & \text{if } x \in [0, a), \\ \frac{2a}{1-2a} & \text{if } x \in [a, \frac{1}{2}), \\ \frac{-2a}{1-2a} & \text{if } x \in [\frac{1}{2}, 1-a), \\ \frac{-(1-a)}{a} & \text{if } x \in [1-a, 1]. \end{cases} \quad (14)$$

The eigenvalues of the Jacobian matrix are given by

$$\mu_1 = c, \quad (15)$$

$$\mu_2 = c - d, \quad (16)$$

with corresponding eigenvectors

$$e_1 = \begin{pmatrix} 1 \\ 1 \end{pmatrix}, \qquad e_2 = \begin{pmatrix} -\delta \\ \varepsilon \end{pmatrix}. \quad (17)$$

The first eigenvector lies in the synchronization subspace $\mathcal{S}$, whereas the second eigenvector is the transversal one. It should be remarked that, thanks to its being piecewise-linear, the coupled map (11) has out of four different results for the eigenvalues, each of them at a different interval, according to the value of $c$ given by Eq. (14). Since the elements of the Jacobian matrix are constant, their eigenvalues/eigenvectors do not depend explicitely on the value of $x$ or $y$ and thus take on the same value for the fixed points $P_1 = 0$ and $P_2 = 1 - a$ in the synchronized state, belonging to intervals 1 and 4 of $[0, 1]$ and identified by $c = (1 - a)/a$ and $c = -(1 - a)/a$, respectively.

In Figure 3(a) we plot the logarithm of the transversal eigenvalue ($|\mu_2|$) as a function of the coupling parameter $d$ for the fixed points $P_1$ and $P_2$ of the coupled maps (11). Hence such periodic orbits are transversely stable (unstable) if these numbers are negative (positive). We see that, regardless of the value taken on by $d$, there will be always some fixed point with transversal eigenvalue having modulus greater than the unity, although the eigenvalue modulus of a given fixed point could be less than unity for some values of $d$. For the fixed point $P_1 = 0$ this interval is $d_1 < d < d_2$, where

$$d_{1,2} = \left(\frac{1-a}{a}\right) \mp 1, \quad (18)$$

and for $P_2 = 1 - a$, $-d_2 < d < -d_1$. For $a = 0.45$, as in Fig. 3, these intervals are $I_1 = (-2.23, -0.22)$ and $I_2 = (0.22, 2.23)$ for the fixed points $P_1$ and $P_2$, respectively. We conclude that, no matter what the value of the coupling strength $d$, there will be always some transversely unstable fixed point embedded in the (chaotic) synchronized state.

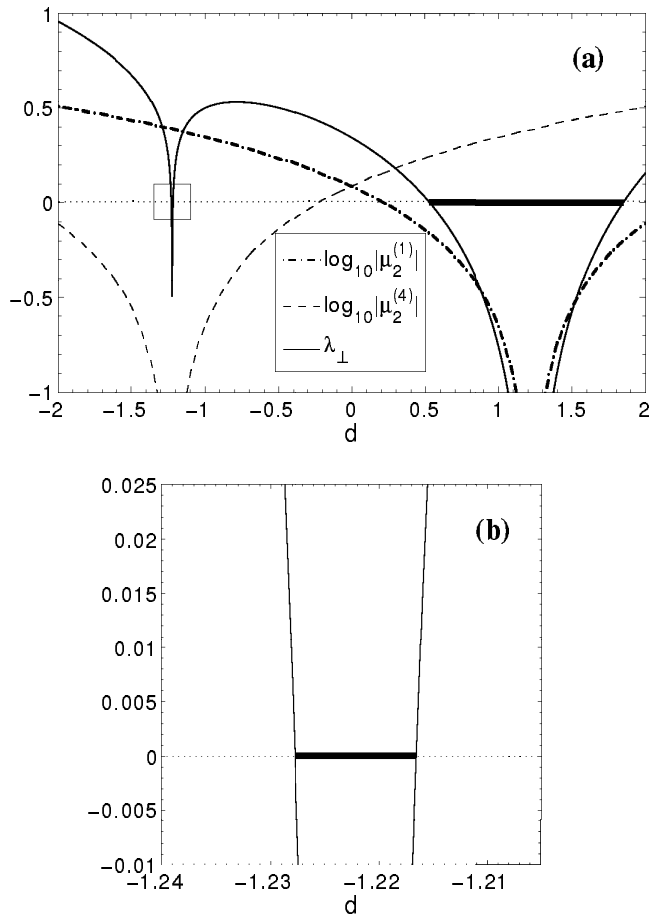

FIG. 3: (a) Logarithm of the transversal eigenvalue and Lyapunov exponent of the coupled piecewise-linear Lorenz maps (version I) as a function of the coupling strength $d = \delta + \varepsilon$ for the fixed points $P_1 = 0$ ($\mu_2^{(1)}$) and $P_2 = 1 - a$ ($\mu_2^{(4)}$). The intervals for riddling are denoted as thicker lines. (b) Magnification of a portion of the diagram.

Note that this analysis has been performed for fixed points only, what it is sufficient for the characterization of riddling in this coupling prescription. Other periodic orbits can be transversely unstable, however, even when some fixed point is transversely stable. Moreover, for the intervals $I_1$ and $I_2$, we observe from Fig. 3(a) that one of the fixed points is transversely stable, whereas the other one is transversely unstable. This phenomenon, called *unstable dimension variability*, has important consequences in terms of the global dynamics of the system [14, 15], such as the validity of computer-generated trajectories of some strongly non-hyperbolic chaotic systems [16] . Moreover, it has applications in chaotic systems with more than one attractor, like the ecological problem of two-species competition with extinction of either one of the species [17]

### B. Version II

Now we turn to the coupled maps in its version II, for which the Jacobian matrix is

$$\mathbf{DG} = c \begin{pmatrix} 1 - \delta & \delta \\ \varepsilon & 1 - \varepsilon \end{pmatrix} \tag{19}$$

where $c$ is given by (14), and whose eigenvectors as well as the eigenvalue $\mu_1$ are the same as in version I, the transversal eigenvalue $\mu_2$ being given by

$$\mu_2 = c(1 - d). \tag{20}$$

The logarithm of the transversal eigenvalue modulus is plotted in Figure 4(a) as a function of the coupling parameter $d$ for the fixed points $P_1$ and $P_2$ of the coupled maps (12). There is a common interval for which both fixed points are simultaneously transversely stable, given by $d_3 < d < d_4$, where

$$d_{3,4} = 1 \mp \left( \frac{a}{1 - a} \right), \tag{21}$$

and which turns to be the interval $I_3 = (0.18, 1.83)$ when $a = 0.45$, as shown in Fig. 4(a). In this case, unlike the previous coupling prescription, it is necessary to analyze also period-2 orbits, since we found that these can be transversely unstable even when the fixed points themselves are transversely stable.

The period-2 points is given by

$$P_1^{(2)} : x_1^{(2)} = \frac{1 - a}{3 - 4a} \in [a, \frac{1}{2}), \tag{22}$$

$$P_2^{(2)} : x_2^{(2)} = f_a(x_1^{(2)}) \in [1 - a, 1], \tag{23}$$

with transversal eigenvalue given by

$$\mu_2^{(per.2)} = \left( \frac{1}{2a - 1} - 1 \right)(1 - d)^2 \tag{24}$$

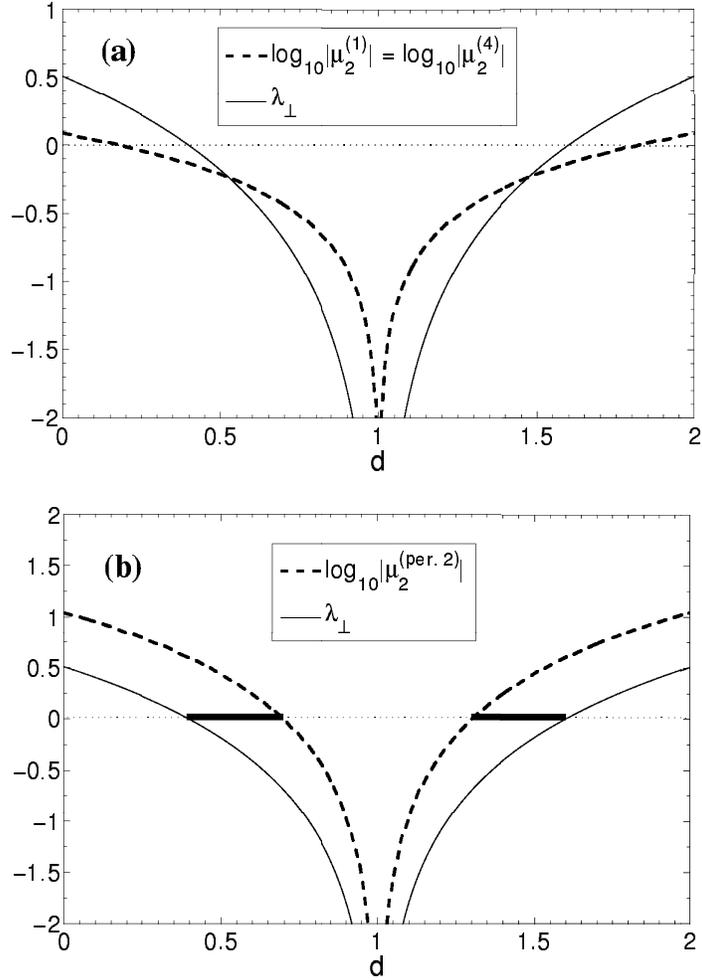

FIG. 4: Logarithm of the transversal eigenvalue and Lyapunov exponent of the coupled piecewise-linear Lorenz maps with $a = 0.45$ (version II) as a function of the coupling strength $d = \delta + \varepsilon$ for (a) fixed points, and (b) period-2 orbit in the synchronization subspace. The intervals for riddling are denoted as thicker lines.

such that the intervals for which these points are transversely unstable are

$$d < 1 - \sqrt{\frac{1-2a}{2(1-a)}}, \quad d > 1 + \sqrt{\frac{1-2a}{2(1-a)}}. \tag{25}$$

The log-eigenvalue of the period-2 orbit is depicted in Fig. 4(b) as a function of $d$, showing that, for the following intervals of $d$: $[0.4, 0.7]$ and $[1.2, 1.5]$, there will be some transversely unstable orbit (either a fixed point or a 2-cycle).

## IV. RIDDLING AND CHAOTIC SYNCHRONIZATION

In this Section we shall describe one of the consequences of the synchronized state being transversely stable as a whole, but possessing transversely unstable periodic orbits embedded on it. For this we focus on the basin of synchronization, which is the set of initial conditions in the phase-space which produce orbits which asymptote to the synchronized state as the time goes to infinity. It may well happen that this basin of synchronization be riddled, i.e. densely punctured with holes (in the measure-theoretical sense) belonging to the basin of another attractor.

The dynamical consequences of riddling are quite severe, for every initial condition we may think of are surrounded by an uncertainty ball of small but nonzero radius, representing the observational errors or parametric noise intrinsic to the determination of a state in the phase space. Hence, if the basin is densely riddled with regions belonging to another basin, such regions will intercept the uncertainty ball, turning uncertain the determination of the final state of the system. Even worse, this problem is not going to diminish reducing the radius of the uncertainty ball, since riddling occurs at every scale, no matter how small.

The conditions for the occurrence of riddled basins in a dynamical system are [18]:

1. There exists an invariant subspace whose dimension be less than of the phase space;

2. The dynamics on the invariant subspace has a chaotic attractor A;

3. There is another attractor B, chaotic or not, and not belonging to the invariant subspace;

4. The attractor A is transversely stable in the phase space, i.e. for typical orbits on the attractor the Lyapunov exponents for infinitesimal perturbations along the directions transversal to the invariant subspace M are all negative;

5. A set of unstable periodic orbits embedded in the chaotic attractor A is transversely unstable.

The fact that the synchronization subspace $\mathcal{S}$ of the coupled maps is invariant under the system dynamics already fulfills condition **1** for riddling. Moreover, if the coupled identical systems undergo chaotic motion, their synchronized state is also chaotic, i.e. there is a chaotic attractor A embedded in the synchronization subspace (condition **2**). For condition

**3** to be satisfied there must exist another attractor B off the synchronized manifold $\mathcal{S}$ (*global riddling*). There are many situations, however, for which there is no such attractor, but the trajectories off-$\mathcal{S}$ spend an arbitrarily large time wandering through the available phase space region before returning to $\mathcal{S}$. In this cases (*local riddling*) we can still think of this non-synchronized transient regime as a second possible behavior and assign to it the properties formerly possessed by the second attractor B, namely conditions **4** and **5** of riddling. Hence, strictly speaking, local riddling means that only the local conditions for riddling have been verified for the system in analysis.

In practice, however, while most of the conditions for riddling rely on a local analysis in the vicinity of the synchronized state, the determination of a second attractor is usually an information of a global nature. Except for some cases in which this can be rigorously proved, the existence of a second attractor is suggested by numerical simulations of the dynamical system. Hence, as far as the local analysis is concerned, we are faced more often with local riddling than global one in complex systems.

A formal way to understand local riddling consists on choosing a linear neighborhood $\Sigma_\Delta$ of the synchronization subspace $\mathcal{S}$, where $\Delta \to 0$, which is represented by a strip of width $2\Delta$ surrounding it in both sides. If we denote with white pixels those initial conditions generating orbits that escape rapidly from $\Sigma_\Delta$ and by black points those orbits which do not escape and are attracted to the synchronization subspace, we can have the same tongue-like basin structure of the globally riddled case, but now restricted to the linear neighborhood $\Sigma_\Delta$.

In the case of global riddling, it should be mentioned that condition **4** for riddling implies that "most" points, in the traditional two-dimensional Lebesgue measure sense, stay within this $\Sigma_\Delta$ neighborhood and are eventually attracted to A, as proved in Ref. [10]. This is thus compatible with the existence of a positive Lebesgue measure set of orbits in the linear neighborhood $\Sigma_\Delta$ that move away from it.

### A.  Version I

In order to verify condition **4** for riddling, we have to investigate for which values of the coupling coefficient $d = \epsilon + \delta$ the transversal Lyapunov exponent is negative. Let us first consider version I of the coupled maps. The Lyapunov exponents along the direction parallel to the synchronization subspace is the same as that for an uncoupled piecewise-linear Lorenz map, Eq. (7). The exponent along the transversal direction, on the other hand, can be obtained by applying Birkhoff ergodic theorem as

$$\begin{aligned}\lambda_\perp &= \lim_{m\to\infty} \frac{1}{m} \sum_{n=0}^{m-1} \ln|\mu_2(x_n)| = \int_0^1 \ln|\mu_2(x)|\,\rho(x)\,dx \\ &= \frac{1}{2-3a}\left[a\ln\left|\frac{1-a}{a}-d\right| + \frac{1-2a}{2}\ln\left|\frac{2a}{1-2a}-d\right| + \right. \\ &\quad \left. \frac{1-2a}{2}\ln\left|\frac{-2a}{1-2a}-d\right|\right] + \frac{1-2a}{2-3a}\ln\left|\frac{-(1-a)}{a}-d\right|, \end{aligned} \quad (26)$$

where we have used the invariant density given by Eq. (6).

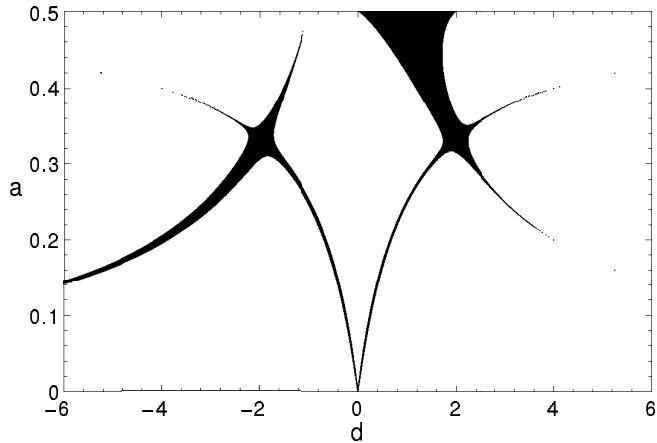

FIG. 5: In the parameter plane $a$ *versus* $d$, the black regions have negative transversal Lyapunov exponent the coupled piecewise-linear Lorenz maps for $a = 0.45$ (version I).

In Fig. 3(a) we plot the transversal Lyapunov exponent as a function of the coupling strength sum $d = \delta + \varepsilon$ for $a = 0.45$. There are two intervals of $d$ for which the transversal Lyapunov exponent is negative, thus fulfilling condition 4 for riddling. The boundaries of these intervals are given by the values of $d$ for which $\lambda_\perp = 0$, yielding transcendental equations in view of (26). The regions, in the parameter plane $a \times d$, characterized by negative transversal exponents, are depicted in Fig. 5. In the piecewise-linear Lorenz case ($a = 0.45$) such intervals are $(-1.224, -1.220)$ and $(0.689, 1.707)$.

In the previous Section we concluded that, in version I of the coupled maps, there will be always transversely unstable fixed points in the synchronization subspace [see Fig. 3]. Besides the fixed point itself being transversely unstable, so are all its pre-images, which form a denumerable infinite set of points, thus condition **5** for riddling is always fulfilled for any value of $d$, such that the only remaining condition to be verified is the negativeness of $\lambda_\perp$, the intervals being thus indicated by Fig. 5 as black regions.

However, we must remark that condition **5** is of a local nature, whereas the existence of a second attractor is an information from the global dynamics of the coupled system. Unless we can rigorously prove that there is a second attractor at infinity (or elsewhere) for a basin with a nonzero Lebesgue measure the set of mathematical conditions can be used to define local riddling. For version I, this proof can be given by writing the coupled maps in terms of the variables $u = x - y$ and $v = x + y$. Since the map is defined in four intervals, so as the coupled maps in version I. Let us consider the interval $0 < x < a$. The dynamics along the transversal direction is piecewise linear, with a slope given by (14). Hence, if $d < d_{C2} = (1 - a)/a$ the unstable manifold of the unstable fixed point extends to infinity, and the pre-images of the fixed point as well. A trajectory that closely follows this unstable manifold will asymptote to infinity, which is the second attractor characterizing global riddling. Moreover, this slope must be positive, otherwise the trajectories would approach the line $x = y$ but asymptote to infinity as well. This leads to another condition, namely $d > d_{C1} = (1 - 2a)/a$.

For global riddling the basin of synchronization is the set of initial conditions which asymptote to the synchronized attractor, with a similar definition for the basin of infinity. If we have local riddling, however, it is not possible to define such basins, hence it is more convenient to consider the synchronization time $\tau$, i.e. the time it takes for an initial condition $(x_0, y_0)$ to reach the synchronized state (i.e. to approach its vicinity with an accuracy of $\sim 10^{-5}$.

In Figure 6 we plot in a gray-scale the values of the synchronization time $\tau$ for the initial condition $(x_0, y_0)$ placed at the center of the corresponding pixel, for a value of $d$ which fulfills the local requirements for riddling. The darker pixels indicate initial conditions whose orbits achieve the synchronized state in less than ten iterations, whereas for the lighter pixels this time is higher. White pixels indicate orbits which go to infinity.

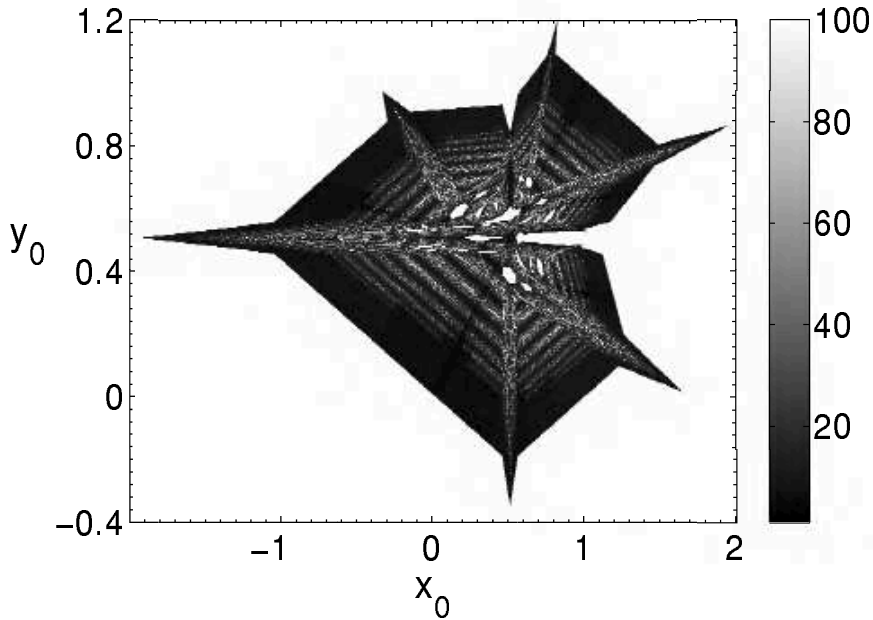

FIG. 6: Synchronization time for initial conditions in the phase plane for $a = 0.45$ and $d = 1.0$ (version I).

## B. Version II

Now for the version II, the transversal Lyapunov exponent is given by

$$\begin{aligned}\lambda_\perp &= \frac{1}{2 - 3a}\left[(1 - a)\ln\left(\frac{1 - a}{a}\right) + (1 - 2a)\ln\left(\frac{2a}{1 - 2a}\right)\right] + \ln|1 - d| \\ &\equiv K(a) + \ln|1 - d|.\end{aligned} \quad (27)$$

where $K$ is a constant for a given $a$. Figure 4 displays the dependence of $\lambda_\perp$ with $d$ for $a = 0.45$, showing an interval $(d_1, d_2) = (0.442, 1.558)$ for which the transversal Lyapunov exponent is negative, thus verifying condition 4 for riddling. In fact, the boundaries of this interval are the roots of $\lambda_\perp = 0$, namely

$$\tilde{d}_{1,2} = 1 \mp e^{-K(a)}, \quad (28)$$

straddling the point $d = 1$ at which the Lyapunov exponent is minus infinity.

We have seen in the previous Section that, for the intervals $[0.4, 0.7]$ and $[1.2, 1.5]$, we can warrant the existence of transversely unstable periodic orbits embedded in the chaotic synchronized attractor, either fixed points or period-2 points, so as to fulfill condition **5** for riddling. The union of these intervals, on its hand, contain the interval $(d_1, d_2)$ which verifies condition **4** for riddling, and hence the latter ultimately determines the existence of riddling. The regions with $\lambda_\perp < 0$ are indicated by black pixels in Fig. 7 as a function of both $a$ and $d$. If, besides having $\lambda_\perp < 0$, the system also presents riddling, we indicate by gray pixels this situation. We observe that, for $a = 1/3$ (i.e. the tent map case), there is no riddling interval, since the system is hyperbolic and $\lambda_\perp < 0$ always implies synchronization [19]. For $a < 1/3$, riddling can be explained from the transversal instability of the fixed points, whereas for $a > 1/3$ we have to analyze a period-2 orbit to verify condition **5** for riddling.

In version II of coupling, it turns out that the unit square $[0, 1]^2$ is kept invariant under the dynamics of the coupled system if both $\epsilon$ and $\delta$ belong to the interval $[0, 1]$. These constraints being respected, there is no attractor in the phase place other than the synchronized one, and riddling (if any) is certainly of a local nature. However, if either $\epsilon$ or $\delta$ take on values outside the interval $[0, 1]$, the unit square is no longer invariant and orbits can, in principle, asymptote to other attractor. We have been able to find numerically a second attractor at infinity, hence we can actually have global riddling.

Following the same procedure used in version I, we also plot in Fig. 8 the synchronization time for initial conditions and a value of $d$ which is inside the riddling interval. The darker pixels, representing "fast" initial conditions, belong to the basin of the synchronized attractor, whereas the brighter pixels represent "slow" initial conditions that either spend a long time until asymptoting to the synchronized attractor or they may escape to the attractor at infinity. The example of riddling given in Fig. 8 shows the regions of "slow" and "fast" initial conditions densely intermixed in fine scales. Moreover, the bands of "slow" initial conditions have the characteristic tongue-like shape that we find in riddling bifurcation scenarios [20].

## V. CONCLUSIONS

The Lorenz equations are a paradigmatic example of deterministic chaos, which has stimulated research over the last four decades. One noteworthy feature of coupled Lorenz equations is the presumed existence of riddled basins for some values of the coupling strength between them, in the sense that the basin structure can be so involved that near a point belonging to some basin there exists a point belonging to other basin, *no matter how near they are*. This is clearly a troublesome feature since any initial condition can be assigned with a small but nonetheless finite uncertainty, and the latter can be so large as to turn the final state of the system completely uncertain if the basins are riddled. This phenomenon is difficult to analyze in the original Lorenz equations due to the high dimensionality, and thus we have considered a low-dimensional model which may shed some light on the dynamical behavior of the high-dimensional system, particularly on the existence of riddled basins.

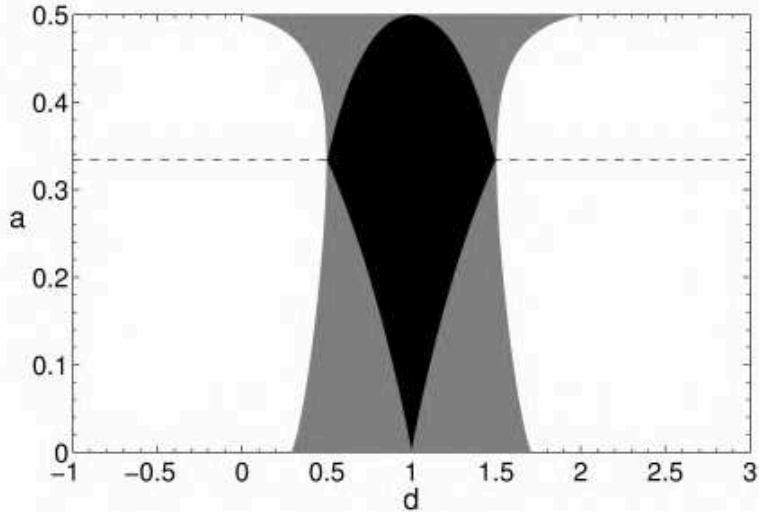

FIG. 7: In the parameter plane *a versus d* the black regions have negative transversal Lyapunov exponent, whereas the gray regions have in addition a transversely unstable orbit, thus characterizing riddling. We considered two coupled piecewise-linear Lorenz maps for $a = 0.45$ (version II).

In this spirit, we analyzed two schemes of coupling piecewise-linear versions of the Lorenz map, since exact analytical results can be given for them, e.g. the natural measure of chaotic orbits and the corresponding Lyapunov exponents. We found that, as a general trend, riddled basins are possible for relatively wide intervals of the coupling strength for both versions, what gives us more confidence that the suggestive evidences of riddling in the original coupled Lorenz systems are in fact true. There are some differences between both schemes, though.

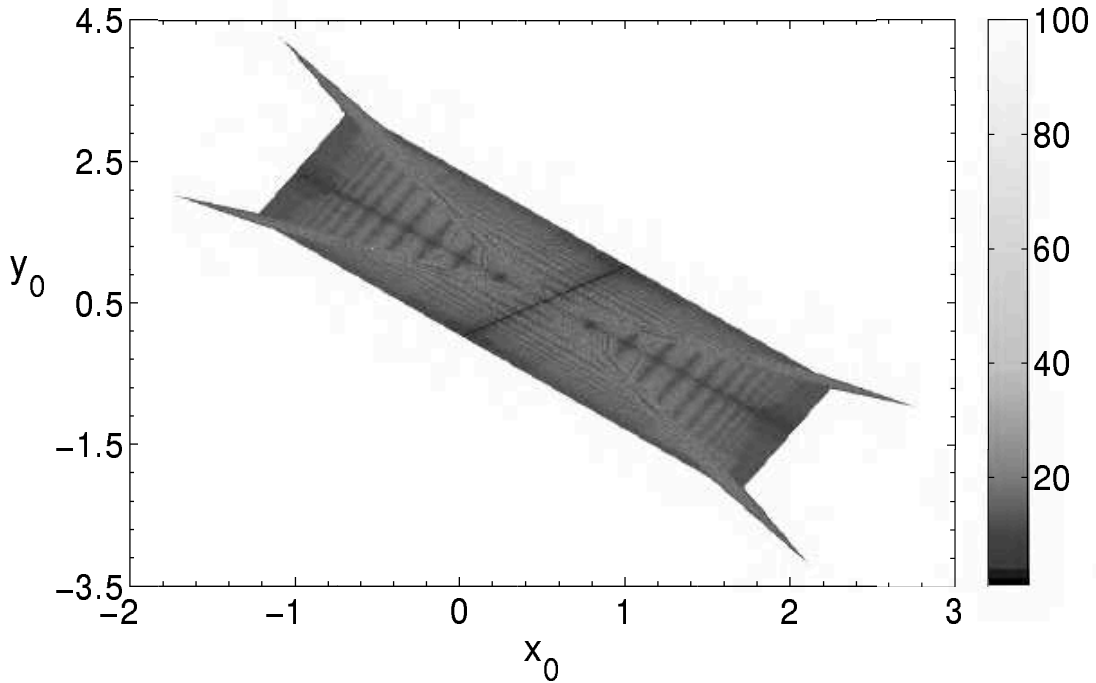

FIG. 8: Synchronization time for initial conditions in the phase plane for $a = 0.45$ and $d = 1.1$ (version II).

When we add the coupling term after applying the map function (version I), the dynamics do not keep invariant the unit square. In this case we found that either the synchronized state has a (globally) riddled structure or it is not transversely stable and it is not an attractor itself. If we couple the variables before applying the map function (version II), the dynamics maintains the unit square invariant for some values of the coupling constants. In those cases we are sure that there cannot exist a second attractor, yielding local riddling. For other values, however, we have not been able to detect this second attractor as well. Unlike the previous version of coupling, we found that there are some intervals for which there are transversely unstable fixed points and/or period-two orbits. Hence we can have presence or absence of local riddling, depending on the parameter values.

We expect that for coupled Lorenz systems, one would observe a scenario more akin to version II, since such simulations typically adopt a diffusive type of coupling. Moreover, the more complex dynamical behavior of Lorenz equations makes it possible to assign one or more coexisting attractors, apart from the synchronized state. While bounded attractors would be less often probable, the existence of an attractor at infinity is quite common, such as in coupled Chua circuits. In those cases our analysis could be extended to the characterization of global riddling, as we have done in Version I.

**Acknowledgments**

This work was partially supported by CNPq, CAPES, and Fundação Araucária (Brazilian agencies).

---

[1] E. Lorenz, J. Atmospheric Sci. **20** (1963) 130.

[2] D. Gulick, *Encounters with Chaos* (McGraw Hill, New York, 1992)

[3] W.-H. Steeb, M. A. van Wyk, and R. Stoop, Int. J. Theor. Phys. **37** (1998) 2653.

[4] R. Stoop and W.-H. Steeb, Phys. Rev. E **55** (1997) 7763.

[5] S. Banerjee, J. A. Yorke, and C. Grebogi, Phys. Rev. Lett. **80** (1998) 3049

[6] E. E. Pavlovskaia and M. Wiercigroch, J. Sound and Vibrat. **305** (2007) 750.

[7] C. Anteneodo, S. E. de S. Pinto, A. M. Batista, and R. L. Viana, Phys. Rev. E **68** (2003) 045202 (R).

[8] M. Hasler and Yu. L. Maistrenko, IEEE Trans. Circ. Syst, I **44** (1997) 856.

[9] Yu. L. Maistrenko, T. Kapitaniak, and P. Szuminski, Phys. Rev. E **56** (1997) 6393. T. Kapitaniak and Yu. L. Maistrenko, Physica D **126** (1999) 18.

[10] J. C. Alexander, J. A. Yorke, Z. You, and I. Kan, Int. J. Bifurcat. Chaos **2** (1992) 795

[11] J. Aguirre, R. L. Viana, and M. A. F. Sanjun, Fractal structures in nonlinear dynamics, Rev. Mod. Phys. (2008) to appear

[12] C.-M. Kim, S. Rim, W.-H. Kye, J.-W. Ryu and Y.-J. Park, Phys. Lett. A **320** (2003) 39.

[13] H. Fujisaka and T. Yamada, Prog. Theor. Phys. **69** (1983) 32; L. M. Pecora and T. L. Carroll, Phys. Rev. Lett. **64** (1990) 821.

[14] E. J. Kostelich, I. Kan, C. Grebogi, E. Ott, and J. A. Yorke, Physica D **109** (1997) 81.

[15] R. F. Pereira, S. E. de S. Pinto, R. L. Viana, S. R. Lopes, and C. Grebogi, Chaos **17**, 023131 (2007).

[16] R. L. Viana, C. Grebogi, S. E. de S. Pinto, S. R. Lopes, A. M. Batista, and J. Kurths, Phys. Rev. E **68** (2003) 067204 .

[17] R. F. Pereira, S. Camargo, S. E. de S. Pinto, S. R. Lopes, and R. L. Viana, Phys. Rev. E **78** (2008) 056214.

[18] E. Ott, J. C. Sommerer, J. C. Alexander, I. Kan, and J. A. Yorke, Phys. Rev. Lett. **71** (1993) 4134. E. Ott, J. C. Alexander, I. Kan, J. C. Sommerer, and J. A. Yorke, Physica D **76** (1994) 384.

[19] R. F. Pereira, S. E. de Souza Pinto, R. L. Viana, and S. R. Lopes, Arxiv Preprint arXiv:0809.0294, 2008 [nlin.CD]

[20] Y.-C. Lai, C. Grebogi, J. A. Yorke, and S. Venkataramani, Phys. Rev. Lett.**77**, 55 (1996).